\documentstyle[12pt]{article}
\begin{document}

\vspace{.5cm}

\centerline {\LARGE Non--Lagrangian Construction}

\centerline {\LARGE of Hamiltonian Structures}

\vspace{1.5cm}

\centerline {\Large Sergio A. Hojman}

\vspace{.4cm}

\centerline {\Large Departamento de F\'{\i}sica}

\centerline {\Large Facultad de Ciencias}

\centerline {\Large Universidad de Chile}

\centerline {\Large Casilla 653, Santiago, CHILE}

\newpage

\baselineskip=1.0cm

\centerline {\Large ABSTRACT}

\vspace{1.0cm}

A method to construct Hamiltonian theories for systems of both ordinary and
partial differential equations is presented. The knowledge of a Lagrangian is
not at all necessary to achieve the result. The only ingredients required for
the construction are one solution of the symmetry (perturbation) equation and
one constant of the motion of the original system. It turns out that the
Poisson bracket structure for the dynamical variables is far from being
uniquely determined by the differential equations of motion. Examples in
classical mechanics as well as in field theory are presented.\\

\newpage

Hamiltonian methods are widely used in connection with problems in classical,
quantum and statistical mechanics, fluid dynamics, optics, solid state,
molecular, atomic, nuclear, particle and plasma physics, in both classical
and quantum field theoretical systems. Quantization schemes as well as group
theoretical symmetry methods are examples of subject matters in which
Hamiltonian structures are useful.  Hamiltonian theories are usually
constructed starting from the knowledge of a Lagrangian, by well established
methods for both the cases of regular and singular Lagrangians
\cite{gold,l-l,dirac,suda}.\\ For different reasons, one may try to quantize
or to construct Hamiltonian structures for classical systems of differential
equations, without recourse to a Lagrangian \cite {wigner,dyson,sah-lcs}.
Several authors have been successful in creating Hamiltonian theories from
scratch for different examples, mostly in fluid dynamics (an excellent review
is presented in \cite{morri}) and in field theory \cite{rebbi}, but {\it no
general method} seems to exist for constructing a Hamiltonian structure
starting from the equations of motion only, without using at all either the
explicit form or the existence of a Lagrangian formulation for the system at
hand.\\ The purpose of this note is to present a {\it general technique} to
construct Hamiltonian theories starting from the equations of motion, using
one symmetry transformation and one constant of the motion, without recourse
to the Lagrangian of the system of equations, which may even fail to exist. A
completely different approach has been used to construct a quantum model for
a non--Lagrangian cosmological model in \cite{sh-mr}), and infinitely many
Hamiltonian structures of the spinning top in \cite{sh1}.  Let us first
define what it is usually meant by a Hamiltonian theory.\\
Consider an autonomous first order differential system,
\begin{equation}
\frac{dx^a}{dt} = f^a(x^b)\ \,\ \ \ \ \ \ \ \ \ \ a,b = 1,........,N\ .
\label{eqmo}
\end{equation}
Of course, a differential system of any order can be easily cast in first
order form by defining extra variables in the standard textbook fashion.
A Hamiltonian structure for (\ref{eqmo}) is defined in terms of an
antisymmetric symplectic matrix $J^{ab}(x^c)$ and a Hamiltonian $H(x^c)$
which satisfy
\begin{equation}
J^{ab} = - J^{ba}  \ \ \ \ \ \ \ \ \ \ a,b,c,..... = 1,........,N\ ,
\label{anti}
\end{equation}
\begin{equation}
{J^{ab}}_{,d}\ J^{dc} + {J^{bc}}_{,d}\ J^{da} + {J^{ca}}_{,d} \ J^{db}\ \equiv
0\ ,
\label{ji}
\end{equation}
and,
\begin{equation}
J^{ab}\  \frac{\partial H}{\partial x^b} = f^a\ .
\label{hami}
\end{equation}
The Poisson bracket between any two dynamical variables $A(x^a)$ and $B(x^b)$
is defined by
\begin{equation}
[A,B]\ =\ \frac{\partial A}{\partial x^a}\ J^{ab}\ \frac{\partial B}{\partial
x^b}\ ,
\label{pb}
\end{equation}
and it satisfies all the usual algebraic and differential properties as it can
be easily inferred from the antisymmetry condition (\ref{anti}),
the Jacobi identity (\ref{ji}), and the definition (\ref{pb}).\\
If, in addition,
\begin{equation}
\det J^{ab}\  \neq 0\ ,
\label{regu}
\end{equation}
is required, then the symplectic matrix is regular. It is important to
remark that, sometimes, condition (\ref{regu}) cannot be met. If $N$ is odd,
as it happens for Euler's equations, (\ref{regu}) is never
satisfied because of (\ref{anti}) as it can be seen, for instance in \cite
{sh1} and in one of the examples below. In Dirac theory, condition
(\ref{regu}) for Dirac brackets is not satisfied because there are dynamical
entities, called {\it second class constraints}, which have vanishing Dirac
brackets with any variable
\cite{dirac}.  In fluid dynamics, the functions which have vanishing Poisson
brackets with any dynamical variable are called ``Casimir functions'',
inspired on the well known group theoretical terminology for operators which
commute with any element of the group
\cite{morri,little}.
Therefore, it
is convenient to adopt a flexible attitude regarding condition (\ref{regu}),
and take conditions (\ref{anti}), (\ref{ji}), and (\ref{hami}) as defining a
Hamiltonian theory. Of course, the usual textbook Hamiltonian structures
satisfy all of them.\\ It is straightforward to prove that conditions
(\ref{anti}) and (\ref{hami}), imply that $H(x^a)$ is a time independent
constant of the motion of system (\ref{eqmo}) defined by the condition
\begin{equation}
\frac{\partial H}{\partial x^a}\ \ f^a\ =\ {\cal L}_f\ H\ \ =\ 0\ \ ,
\label{const}
\end{equation}
which can be equivalently stated by saying that the Lie derivative of $H$
along $f$ vanishes. A brief account on Lie derivatives may be found in
\cite{wein}.\\
Let us now derive the symmetry (perturbation) equation of (\ref{eqmo}). (For a
detailed discussion, see \cite{sh2}).
Consider the transformation
\begin{equation}
{\tilde{x}}^a\ =\  x^a\ +\ \epsilon \ \eta^a(x^b,t)\ \ ,
\label{pert}
\end{equation}
where $\epsilon \ \eta^a(x^b,t)$ is a small perturbation which maps solutions
of (\ref{eqmo}) in solutions of the same equation, up to first order in
$\epsilon$. The equation that the perturbation vector $\eta$ satisfies is,
\begin{equation}
\partial_t\ \eta^a\ +\ {\eta^a}_{,b}\ \ f^b\ -\ {f^a}_{,b}\
\eta^b \ =\ 0\ \ \ ,
\label{symm}
\end{equation}
or,
\begin{equation}
(\partial_t\ \ +\ \ {\cal L}_f)\ \eta^a\ \ =\ 0\ .
\label{symmlie}
\end{equation}
It is not difficult to prove that $K$, the deformation of $H$ along $\eta$,
defined by
\begin{equation}
K\ \equiv\ \frac{\partial H}{\partial x^a}\ \ \eta^a\ =\ {\cal L}_\eta\ H\
\ \ ,
\label{deform}
\end{equation}
is also a constant of the motion for the same system, if $\eta$ satisfies Eq.
(\ref{symmlie}). By the same token, a new symmetry transformation
$\bar{\eta}$ which satisfies Eq. (\ref{symmlie}) can be constructed using
a symmetry transformation $\eta$ and a constant of motion $K$ by
\begin{equation}
{\bar{\eta}}^a\ =\ \frac{\eta^a}{K}\ \ .
\label{newsymm}
\end{equation}
A detailed account of these results may be found in \cite{sh3}.\\
Let us now compute the Lie derivative of $J^{ab}$ along $f$
\begin{equation}
{\cal L}_f \ J^{ab} \ = \ {J^{ab}}_{,c} \ f^c \ - \ J^{ac} \ {f^b}_{,c}\ -
\ J^{cb} \ {f^a}_{,c} \ \ \ .
\label{liej}
\end{equation}
It is an straightforward exercise to prove that
\begin{equation}
{\cal L}_f \ J^{ab} \ = \ 0 \ \ \ .
\label{liej0}
\end{equation}
using Eqs. (\ref{anti}), (\ref{ji}), and (\ref{hami}). Note that
Eq. (\ref{regu}) is not needed in the proof. Therefore, a symplectic matrix
must have vanishing Lie derivative along $f$. Nevertheless, this condition is
not sufficient to fullfill simultaneously the requirements (\ref{anti}),
(\ref{ji}), and (\ref{hami}). To construct a symplectic matrix $J^{ab}$, let
us start by considering an antisymmetric matrix according to the following
{\it ansatz}
\begin{equation}
J^{ab}\  =\  f^a\ \eta^b\ -\ f^b\ \eta^a\  ,
\label{ans}
\end{equation}
where $\eta$ satisfies (\ref{symm}) and has been normalized using
(\ref{newsymm}) in such a way that $J^{ab}$ fulfill (\ref{hami}) identically.
Of course, condition (\ref{anti}) is trivially met. The Jacobi identity
(\ref{ji}) imposes the following condition
\begin{equation}
J^{bc}\ {\cal L}_f\ \eta^a\   +\  J^{ca}\ {\cal L}_f\ \eta^b\   +
\ J^{ab}\ {\cal L}_f\ \eta^c\   =\ 0\ ,
\label{jian}
\end{equation}
which is satisfied by a particular, time independent symmetry vector $\eta_0$
which solves (\ref{symmlie}) defined by
\begin{equation}
\partial_t \ {\eta_0}^a\ =\ - \ {\cal L}_f\ {\eta_0}^a\ \ =\ 0\ .
\label{symm0}
\end{equation}
A more interesting solution $\eta_1$ is given by the condition
\begin{equation}
\partial_t \ {\eta_1}^a\ =\ - \ {\cal L}_f\ {\eta_1}^a\ \ =\ \lambda \ f^a\ ,
\label{symm1}
\end{equation}
which will be most useful in many instances. Note that both solutions produce
symplectic matrices with vanishing Lie derivatives along $f$.\\
We have have thus contructed a Hamiltonian structure for (\ref{eqmo}) based
on the knowledge of just one symmetry vector (either $\eta_0$ or $\eta_1$)
and only one constant of the motion, $H$, of the system under consideration
(assuming a non vanishing $K$, which can be easily achieved as it will be
seen in the examples).\\
A few  comments seem in order. First, a solution similar to the one described
by (\ref{symm1}), with the Lie derivative of the symmetry vector along $f$
proportional to the symmetry vector itself, although it satisfies the Jacobi
identity, is incompatible with (\ref{hami}). Second, the rank of the symplectic
matrix just derived is two. Therefore, it will be, in most cases, singular. A
procedure to enlarge its rank will be described below. Third, it is obvious
that the method we have presented will, in general, yield a Hamiltonian
structure written in terms of non--canonical coordinates. Nevertheless, that
is the most we can hope for in the case of non--Lagrangian systems (which are
always described by a non--commutative geometry). This problem is dealt with
in some detail in \cite{sah-lcs}. Fourth, even though this procedure differs
from the usual one for the case of Lagrangian systems, it may sometimes
reproduce the well known results in terms of canonical coordinates, as it is
shown in one of the examples below.\\ Let us now consider some examples. The
systems may be completely described by the evolution vector $f$, or the
equations of motion written in its first order version. The Hamiltonian
structure may be completely determined by one constant of motion, the
Hamiltonian $H$, and one symmetry vector $\eta_0$ or $\eta_1$. Sometimes, we
will need to make use of the the normalization given in (\ref{newsymm}).\\

{\bf Example 1.--} One dimensional monomial force.\\
This example is defined by the equations of motion
\begin{equation}
f^1\ =\ x^2\ ,\ f^2\ =\ -\ c\ (n+1)\ (x^1)^n\ ,
\label{e1}
\end{equation}
while a Hamiltonian
\begin{equation}
H\ =\ \frac{(x^2)^2}{2}\ +\ c\ (x^1)^{n+1}\ ,
\label{e2}
\end{equation}
and one symmetry transformation are given by
\begin{equation}
\eta^1\ =\ x^1\ +\ \frac{n-1}{2}\ t\ f^1\ ,\ \eta^2\ =\
\frac{n+1}{2}\ x^2\ + \frac{n-1}{2}\ t\ f^2\ .
\label{e02}
\end{equation}
This is, of course, a very trivial example, which, nonetheless, shows how the
scheme presented here can reproduce the usual results. In this case, the
symplectic matrix is regular. The harmonic oscillator and the free particle
are special cases in this example. Note that this treatment can be extended to
any number of dimensions provided the force be a homogeneous function of
degree $n$ in the coordinates.\\

{\bf Example 2.--} Euler's Top.\\
Consider the equations of motion of Euler's top
\begin{equation}
\frac{dL^i}{dt} = - \epsilon^{ijk} \Omega_j L_k\ \equiv f^i \ \ \,\ \ \ \ \
\ \ \ \ \ i,j,k = 1,2,3 \ \ \ ,
\label{eqtop}
\end{equation}
with
\begin{equation}
\Omega_j = \frac{L_j}{I_j}  \ \ \ ,
\label{defom}
\end{equation}
where $L_i = L^i$ and $\Omega_i = \Omega^i$  are the components of the
angular momentum vector and the angular velocity vector in the $\rm
i^{th}$ principal direction respectively and the $I_i$ is the eigenvalue of
the tensor of inertia of an asymmetrical top along the $\rm i^{th}$ principal
axis, as usual.\\
Let us now look for symmetries of the equations of motion. With this purpose
in mind, multiply the angular momentum by some constant factor $\lambda$.
This operation introduces a $\lambda^2$ factor in the right hand side of the
equation of motion (\ref{eqtop}). The same result is achieved in the left hand
side of the equation if, in addition, time is multiplied by the inverse
factor $\lambda^{-1}$. These operations performed simultaneously constitute a
finite symmetry transformation for (\ref{eqtop}). One can deal with an
infinitesimal version of it by considering $\lambda = 1 +
\zeta$ infinitesimally close to one, to get the transformation
\begin{equation}
\delta L^i = \zeta L^i , \ \ \ \ \delta t = - \zeta t \ \ ,
\label{dldt}
\end{equation}
Note that a transformation such as (\ref{dldt}) may
equivalently be written as (see, for instance \cite{sh2})
\begin{equation}
\eta^i = \zeta (L^i +  t  \epsilon^{ijk} \Omega_j L_k)\ \ ,
\label{etal}
\end{equation}
leaving time invariant. It is now a straightforward matter to check that the
transformation defined by (\ref{etal}) is, in fact, a symmetry transformation
for (\ref{eqtop}) because it satisfies (\ref{symm}). Note that $\eta^i$ also
satisfies (\ref{symm1}), and consequently may, in principle, be used to define
a symplectic matrix according to (\ref{ans}).\\
It is well known that $C_1$ and $C_2$ given by
\begin{equation}
C_1 = {(L^1)}^2 +{(L^2)}^2 + {(L^3)}^2  \ \ \ ,
\label{c1}
\end{equation}
and
\begin{equation}
C_2 = \frac{{(L^1)}^2}{2 I_1} +\frac{{(L^2)}^2}{2 I_2}
+\frac{{(L^3)}^2}{2 I_3} \ \ \ ,
\label{c2}
\end{equation}
are constants of the motion for the dynamics generated by (\ref{eqtop}).
We have already seen that the Hamiltonian for any system must be a constant
of the motion. Therefore, $C_1$ and $C_2$ are, in principle, possible
Hamiltonians for the top.\\
The deformations of $C_1$ and $C_2$ along $\eta^i$ do not vanish, in fact,
\begin{equation}
K_1\ \equiv\ \frac{\partial C_1}{\partial L^i}\ \ \eta^i\ =\ 2 C_1\ \ \ ,
\label{k1}
\end{equation}
and
\begin{equation}
K_2\ \equiv\ \frac{\partial C_2}{\partial L^i}\ \ \eta^i\ =\ 2 C_2\ \ \ .
\label{k2}
\end{equation}
We have thus found two inequivalent Hamiltonian formulations for the top,
defined by the symplectic matrices ${J_1}^{ij}$ and ${J_2}^{ij}$
and the Hamiltonians $H_1$ and $H_2$ given by
\begin{equation}
{J_1}^{ij}\  = \frac{1}{K_1}\ \ ( \  f^i\ \eta^j\ -\ f^j\ \eta^i \ )\ \ \  ,
\label{j1top}
\end{equation}
\begin{equation}
H_1\  = C_1\ \ \  ,
\label{h1top}
\end{equation}
\begin{equation}
{J_2}^{ij}\  = \frac{1}{K_2}\ \ ( \  f^i\ \eta^j\ -\ f^j\ \eta^i \ )\ \ \  ,
\label{j2top}
\end{equation}
and
\begin{equation}
H_2\  = C_2\ \ \  .
\label{h2top}
\end{equation}
Note that the choice of a Hamiltonian $H$ as an arbitrary function $H = H (
C_1 , C_2 )$ is also possible provided the proper normalization factor $K$ is
used in the symplectic matrix $J^{ij}$. In this way, we have constructed
infinitely many Hamiltonian structures for Euler's top.\\

{\bf Example 3.--}  Radial forces.\\
This example considers non--potential radial forces defined by
\begin{equation}
f^i\ =\ x^{i+3}\ ,\ f^{i+3}\ =\ F\ x^i\ ,\ F \ = \ F({\vec{r}}^2,
{\dot{\vec{r}}}^2,{\vec{r}}\cdot{\dot{\vec{r}}}),
\ i\ =\ 1,2,3\ .
\label{e5}
\end{equation}
To construct a Hamiltonian structure, the Hamiltonian may be chosen to be
the third component of the (conserved) angular momentum vector, while the
symmetry transformation is, for instance, a rotation around the first axis.
This example clearly shows the ambiguity which exists to choose both the
Hamitonian and the symplectic matrix.\\

{\bf Example 4.--} Korteweg--de Vries equation.\\
The equation of motion is
\begin{equation}
u_t\ = \ - \ u\ u_x\ - \ u_{xxx}\  \equiv \  f\ \ .
\label{e30}
\end{equation}
We are now going to construct a symmetry transformation for it. Take any
solution of Eq. (\ref{e30}) and define a new
set of variables  $u'$, $x'$, and $t'$ by
multiplying the old variables $u$, $x$, $t$ by factors $\lambda^{-2}$,
$\lambda$, $\lambda^3$ respectively. This operation simply produces an overall
$\lambda^5$ factor in the equation, which means that the new set of
variables solves the same equation which the old variables satisfy. We have
thus constructed a finite symmetry transformation for the Korteweg--de Vries
equation. The infinitesimal symmetry associated to it may be written taking
$\lambda = 1 + \zeta$, infinitesimally close to one
\begin{equation}
\delta u \ = \ - 2 \zeta u \ \ , \ \ \ \delta x \ = \ \zeta x \ \ , \ \
\delta t \ = \ 3 \zeta t \ \ \ ,
\label{e31}
\end{equation}
or, equivalently,
\begin{equation}
\eta  \ = \ \zeta \ (\ -\ 2\ u\ -\ x\ u_x\ +\ 3\ t\ (\ u\ u_x\ +\ u_{xxx}\
 )\ ) \ \ \ ,
\label{e32}
\end{equation}
leaving $x$ and $t$ unchanged \cite{sh2}. It is now a straightforward matter to
prove that $\eta$ is an infinitesimal symmetry transformation for (\ref{e30})
because it satisfies the symmetry (perturbation) equation (\ref{symm}).
We remark that $\eta$ satisfies condition (\ref{symm1}) and
therefore, it can be used to construct a symplectic matrix. \\
One possible choice for the Hamiltonian density is $u^2$. The
\mbox{Hamiltonian $H$}
\begin{equation}
H \ = \ \int u^2 dx \ \ \ ,
\label{e34}
\end{equation}
is a constant of motion, i.e., its time derivative vanishes when the usual
assumptions about the behaviour of the fields at spatial infinity are
adopted. In fact, the time derivative of the Hamiltonian can be written as the
integral of a total spatial divergence (a partial derivative with respect to
$x$, in our case) when the equation of motion (\ref{e30}) is taken into
account.\\
One gets that the deformation $K$ of $H$ along $\eta$ is non--vanishing.
We can easily see that the functional derivative of $H$ in the $\eta$
direction is
\begin{equation}
K\ \equiv\ \int \frac{\delta H}{\delta u(x)}\ \ \eta(x)\ dx \ =\ - \ 3 \ \zeta
\ H\ \ \ ,
\label{e35}
\end{equation}
as it can be obtained by direct computation, or by taking advantage of the
of new variables defined by multiplying the old ones by powers of $\lambda$
as we have already done above.\\
Therefore, one symplectic structure is given by
\begin{equation}
J ( x , y )\  = \frac{1}{K}\ \ ( \  f(x)\ \eta(y)\ -\ f(y)\ \eta(x) \ )\ \ \  .
\label{e36}
\end{equation}
The equation of motion can now be written in Hamiltonian form
\begin{equation}
u_t \  = [\  u\  ,\  H\  ]\ \ \  ,
\label{e37}
\end{equation}
where the field theoretical Poisson bracket is defined, as usual, in terms of
functional derivatives by
\begin{equation}
[\ A\ ,\ B\ ] \ \equiv \ \int  \frac{\delta A}{\delta u(x)} \ J ( x , y )\
\frac{\delta B}{\delta u(y)}\ dx\ dy \ \ \  .
\label{e38}
\end{equation}
As far as we know this is a new Hamiltonian structure for the KdV equation.
Note that other Hamiltonian densities $H'$ can be used as well, in conjunction
with the same symmetry vector $\eta$, provided the corresponding deformations
$K'$ be used in the definition of the new symplectic matrix $J'(x,y)$.\\

{\bf Example 5.--} Non--linear Schr\"{o}dinger equations.\\
The equations of motion are
\begin{equation}
i\ \psi_t\ +\ \psi_{xx}\ +\ \psi^2\ \psi^*\ =\ 0\ ,
\label{e7}
\end{equation}
and its complex conjugate. One possible non--standard Hamiltonian density is
$\psi\ \psi^*$, and the symmetry vectors are
\begin{equation}
\eta\ =\ -\ \psi\ -\ x\ \psi_x\ +\ 2\ t\ (\ \psi_{xx}\ +\ \psi^2\ \psi^*\ )\ ,
\label{e8}
\end{equation}
and its complex conjugate. This Hamiltonian structure also appears to be new.\\

Of course, some of these structures, have singular symplectic matrices. One
way to increase the rank of the symplectic matrix, without altering any other
of its properties, is the following. Assume we can find two new time
independent symmetry vectors $\eta_2$ and $\eta_3$ such that the Lie
derivatives of the Hamiltonian $H$ along them vanish, i.e.,
\begin{equation}
\frac{\partial H}{\partial x^a}\ {\eta_2}^a\ =\ \frac{\partial H}{\partial
x^a}\ {\eta_3}^a\ =\ 0\ ,
\label{e9}
\end{equation}
and that the Lie derivatives of $\eta_2$ along $\eta_3$ as well as those of
$\eta_2$ and $\eta_3$ along $\eta_1$ (or $\eta_0$) vanish. Then, the new
symplectic matrix ${J_1}^{ab}$ defined by
\begin{equation}
{J_1}^{ab}\ =\ J^{ab}\ +\ {\eta_2}^a {\eta_3}^b\ -\ {\eta_3}^a {\eta_2}^b\ ,
\label{e10}
\end{equation}
satisfies all of the requirements which define a symplectic matrix
(\ref{anti}), (\ref{hami}), and even the non--linear Jacobi identity
(\ref{ji}), and its rank is equal to four. This procedure can be repeated at
will, producing an increase of two units in the rank of the
symplectic matrix each time that it is performed. (Note that this construction
clearly shows that the Poisson bracket structure is not uniquely determined
by the dynamics). If, eventually, one gets a
regular symplectic matrix, the method presented here may constitute an
alternative to construct a Lagrangian description of the system (\ref{eqmo}),
yielding a novel, symmetry based,  approach to the classical Inverse Problem
of the Calculus of Variations \cite{helm,darboux,douglas,sh-lfu}.\\
Note that the choice of the symmetry vector ($\eta_0$ or $\eta_1$) needed to
define the symplectic matrix is determined solely by the requirement of
getting a non--vanishing $K$, given $H$. We have used in the examples both
time--dependent and time--independent symmetry vectors, to illustrate
different possibilities. Sometimes, there may be several adequate choices
of symmetry vectors for a given $H$, producing different Hamiltonian
formulations for the same system.\\We remark that singular symplectic matrices
are present in Dirac's construction of Hamiltonian structures, as we have
already mentioned above.  We are currently investigating the possibility of
applying this method, which naturally leads to singular symplectic matrices,
to deal with gauge and constrained systems, as an alternative to Dirac's
method when no Lagrangian is available. We are also studying whether it is
possible to obtain constants of the motion of the system at hand as Casimir
functions of the singular symplectic matrix constructed here.\\

\vspace{1.0cm}

\centerline {\Large ACKNOWLEDGMENTS}
The author is deeply indebted to P. Ripa and J. Sheinbaum for inspiration
which lead him to undertake the study of Hamiltonian systems. It is a
pleasure to thank  P.J. Morrison for enlightning discussions. Several
interesting conversations with O. Casta\~nos, A.M. Cetto, A. Frank, A.
Gomberoff, L. de la Pe\~na, M.P. Ryan, Jr., E.C.G. Sudarshan, L.C. Shepley,
L.F. Urrutia, at different times and places, are gratefully acknowledged.
This work has been supported in part by Fondo Nacional de Ciencia y
Tecnolog\'{\i}a (Chile) grant 93--0883, and a bi--national grant funded by
Comisi\'on Nacional de Investigaci\'on Cient\'{\i}fica y
Tecnol\'ogica--Fundaci\'on Andes (Chile) and Consejo Nacional de Ciencia y
Tecnolog\'{\i}a (M\'exico).\\

\end{document}